\documentclass[12pt]{article}

\newcommand{\la}{\lambda}
\newcommand{\fC}{\mathrm{C}}
\newcommand{\fS}{\mathrm{S}}

\usepackage{graphicx}
\usepackage{amsmath}
\usepackage{amssymb}

\begin{document}
\title{Temporal Talbot effect in interference of matter waves from arrays
of Bose-Einstein condensates and transition to Fraunhofer diffraction}

\author{
A. Gammal$^{1}$ and A.M. Kamchatnov$^{2}$\\
$^1$Instituto de F\'{\i}sica, Universidade de S\~{a}o Paulo,\\
05315--970, C.P. 66318 S\~{a}o Paulo, Brazil\\
$^2$Institute of Spectroscopy, Russian Academy of Sciences,\\ Troitsk 142190,
Moscow Region, Russia
}

\maketitle

\begin{abstract}
We consider interference patterns produced by coherent arrays of
Bose-Einstein condensates during their one-dimensional expansion.
Several characteristic pattern structures
are distinguished depending on value of the evolution time.
Transformation of Talbot ``collapse-revival'' behavior to
Fraunhofer interference fringes is studied in detail.
\end{abstract}

\section{Introduction}

The interference measurements \cite{interf,shin} on two expanding Bose-Einstein
condensates (BECs) have created new important field of research where the
density profile of gas, imaged after releasing from the trap, provides important
information about the phase of the ground-state wave function.
Expansion of coherent arrays of BECs provides new opportunities
to test the phase properties of the system
\cite{orzel}--\cite{greiner2}. For example, Fraunhofer
interference patterns observed in \cite{Pedri} demonstrate strong
coherence of BECs confined in separate traps, and experiment
\cite{greiner2} shows that this coherence can be manipulated by
means of collapses and revivals of wave functions due to
nonlinear interaction of BECs in tightly confined
states formed by three-dimensional periodic trapping potential.

It is well known (see, e.g., \cite{BMS}) that mentioned above
``collapse-revival'' behavior of quantum-mechanical wave
functions \cite{ENSM,AP2} is a temporal counterpart of optical
Talbot effect \cite{talbot,patorski} in which interference pattern
behind the grating restores at distances multiple of the so-called
Talbot distance $d^2/\la$ ($d$ is the slit spacing in the grating
and $\la$ is the wavelength of light). Similar Talbot effect was
also observed in atom optics \cite{CL94,CEH95}. Analogy between
spatial Talbot effect and temporal collapse-revival behavior of wave
functions suggests that such collapses-revivals should exist in
interference of matter waves emitted from arrays of BECs provided
evolution time is small enough, and indeed such effect was observed
in \cite{DHD99}. In this connection it is natural to ask how this
short-time Talbot behavior evolves for finite array of condensates
to long-time Fraunhofer behavior observed in \cite{Pedri}.
This paper is devoted to consideration of this problem.

In Section 2 we present general formulas for the wave function produced by
a linear array of BECs. We confine ourselves with one-dimensional
theory under supposition that condensate remains confined in
radial direction after turning off a periodic optical potential and
evolution takes place only along the axial direction of the BECs array.
This formulas permit us to distinguish characteristic stages of
evolution---short-time Talbot stage with revivals of the wave function
in the central part of the array, intermediate time stage when
Fraunhofer fringes already formed with Fresnel diffraction
pattern inside each of them, and long-time Fraunhofer stage
with standard density distribution along fringes. These stages of
evolution of the wave function are studied in detail in Section 3
(Talbot stage) and Section 4 (transition to Fraunhofer stage).
The last Section 5 is devoted to conclusions.

\section{General formulas}

After switching off the periodic optical potential the condensate
density decreases and under condition that the initial size $\sim\sigma$
of each BEC is much less than the spacing $d$ between sites, the
interatomic interaction can be neglected during most time of the
evolution and, hence, the wave function obeys the
linear Schr\"odinger equation
\begin{equation}\label{FREE}
i\hbar\psi_t=-\frac{\hbar^2}{2m}\psi_{xx}.
\end{equation}
If the initial state is given by $\psi(x,0)=\psi_0(x)$, then after time
$t$ it evolves into
\begin{equation}\label{evolution}
    \psi(x,t)=\int_{-\infty}^\infty G(x-x',t)\psi_0(x')dx',
\end{equation}
where $G(x-x',t)$ is well-known Green function of Eq.~(\ref{FREE})
(see, e.g. \cite{FH65}):
\begin{equation}\label{green}
    G(x-x',t)=\sqrt{\frac{m}{2\pi i\hbar t}}\exp\left[\frac{im(x-x')^2}
    {2\hbar t}\right].
\end{equation}
To simplify calculations, we suppose that the initial wave function
of BEC in the site of the array with the coordinate $kd$ can be
approximated by a Gaussian function and, hence, we represent the
initial state of BEC as
\begin{equation}\label{init}
    \psi(x,0)=\frac{1}{\pi^{1/4}\sqrt{\sigma}}\sum_kA_ke^{i\phi_k}\exp\left[
    -\frac{(x-kd)^2}{2\sigma^2}\right],
\end{equation}
where $N_k=|A_k|^2$ is equal to number of atoms in $k$th condensate
(we suppose that $\sigma\ll d$)
and $\phi_k$ is its phase. Then Eq.~(\ref{evolution}) yields the solution
\begin{equation}\label{train-time}
\begin{split}
    \psi(x,t)=\frac{1}{\pi^{1/4}\sqrt{\sigma(1+i\hbar t/m\sigma^2)}}\sum_{k}
    A_ke^{i\phi_k}\exp\left[-\frac{(x-kd)^2}
    {2\sigma^2(1+i\hbar t/m\sigma^2)}\right].
    \end{split}
\end{equation}
This formula should be specified in accordance with the problem under
consideration. In the case of large number of condensates in the array
confined in axial direction by a parabolic potential,
the Thomas-Fermi approximation can be used  for calculation
of number of atoms in $k$th condensate which gives \cite{Pedri}:
\begin{equation}\label{Nk}
    N_k=A_k^2=\frac{15N}{16k_M}\left(1-\frac{k^2}{k_M^2}\right)^2,
\end{equation}
where
\begin{equation}\label{kM}
    k_M=\sqrt{\frac{2\hbar\bar{\omega}}{m\omega_x^2d^2}}\left(
    \frac{15}{8\sqrt{\pi}}N\frac{a}{a_{ho}}\frac{d}{\sigma}\right)^{1/5},
\end{equation}
$N=\sum N_k$ is the total number of atoms, $\bar{\omega}=(\omega_x
\omega_\bot^2)^{1/3}$ is the geometric mean of the magnetic trap
frequencies, $a_{ho}=\sqrt{\hbar/m\bar{\omega}}$ is the corresponding
oscillator length, and $a>0$ is the $s$-wave scattering length.

In the experiment \cite{Pedri} there was $k_M\cong 10^2 \gg 1$,
and this large parameter suggests that there are different stages of evolution.

For
\begin{equation}\label{cr1}
    t\ll\frac{md^2}\hbar
\end{equation}
each condensate evolves independently of each other and there is no
their interference effects.

For
\begin{equation}\label{cr2}
    t\sim\frac{md^2}\hbar\ll k_M\cdot\frac{md^2}\hbar
\end{equation}
we have interference between condensates, but in the central part of the
array the influence of its finite size is negligibly small and
local interference pattern can be approximated by that of an
infinite periodic lattice of condensates which leads to temporal
Talbot effect.

For
\begin{equation}\label{cr3}
     k_M\cdot\frac{md^2}\hbar\ll t \ll  k_M^2\cdot\frac{md^2}\hbar
\end{equation}
the Fraunhofer fringes begin to form. Indeed, their positions are given by
(see, e.g., \cite{Pedri})
\begin{equation}\label{peaks}
    x_n(t)\cong\pm n\frac{2\pi\hbar}{dm}\,t,\quad n=0,\,1,\,2,\ldots,
\end{equation}
and if $t$ satisfies the condition (\ref{cr3}), then distances
between neighboring fringes $\sim{2\pi\hbar t}/md$ are much greater than
the size of each fringe $\sim2k_Md$ (see below). At the same time, the
interference pattern inside each fringe is formed by only some part of
the array and hence we get Fresnel diffraction pattern along the fringe.

At last, for
\begin{equation}\label{cr4}
    t\gg k_M^2\cdot\frac{md^2}{\hbar}
\end{equation}
we arrive at usual Fraunhofer diffraction when
the whole array contributes into interference pattern inside each fringe.

\begin{figure}[ht]
\centerline{\includegraphics[width=10cm,height=10cm,clip]{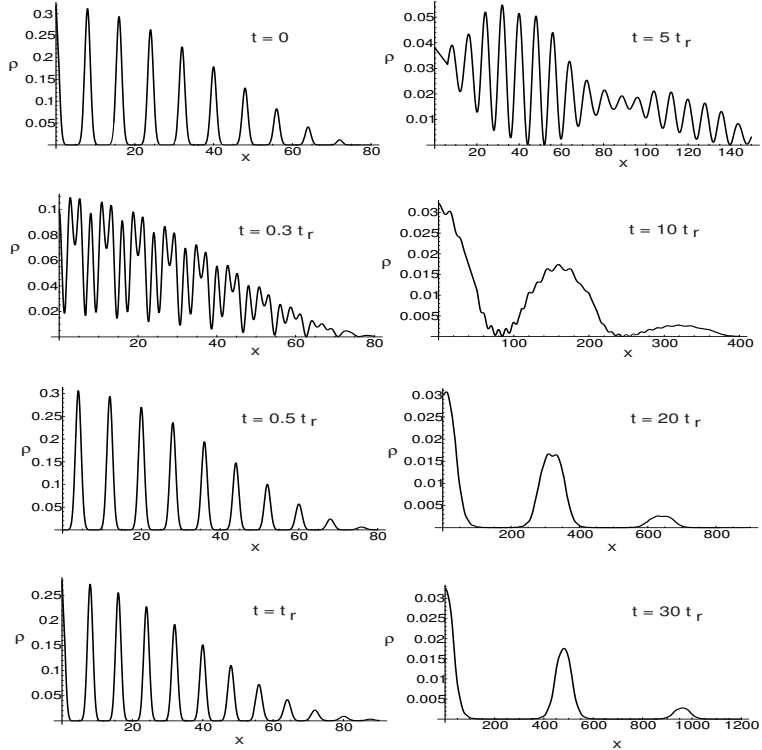}}
\vspace{0.3 true cm}
\caption{
Evolution of the density profile of array of condensates with time
calculated according to Eqs.~(\ref{train-time})--(\ref{kM})  with $d=8$,
$\sigma=1$ (in dimensionless units) and $k_M=10$.
At $t=0.3t_r$, where $t_r$ is given by Eq.~(\ref{period}), we see
complex interference pattern (``collapse''
of wave function); at $t=0.5t_r$ the central
part coincides with that for $t=0$ but shifted to a half-period $d/2$;
at $t=t_r$ the initial distribution is almost completely restored;
at $t=5t_r$ the side fringes start to form, and, finally, at $t=30t_r$ we see
Fraunhofer diffraction of matter waves from finite ``grating''.
}
\label{fig1}
\end{figure}

To illustrate these stages of evolution of the wave function, we have
shown in Fig.~1 the distributions of density $\rho=|\psi|^2$
calculated from formulas (\ref{train-time})--(\ref{kM}) with
$\phi_k=0$ (coherent condensates).
We see that for $t\ll k_M(md^2/\hbar)$ the
density distribution reproduces periodically in time with period
$ t_r\sim md^2/\hbar$ (see exact formula (\ref{period}) below),
the side fringes begin to form at $t\sim k_M(md^2/\hbar)$, and for
$t\gg k_M(md^2/\hbar)$ there are
peaks of density at the coordinates given by (\ref{peaks})
and profiles of fringes take Fraunhofer form for $t\gg (k_Md)^2m/\hbar$.
The solution (\ref{train-time}) permits us to investigate these stages
of evolution analytically.

\section{Talbot revivals of wave function}

For time values in the region (\ref{cr2}),
we can approximate the array by infinite lattice of equidistant
condensates so that for coherent condensates their wave function is given by
\begin{equation}\label{inphase}
\begin{split}
    \psi(x,t)=\frac{A}{\pi^{1/4}\sqrt{\sigma(1+i\hbar t/m\sigma^2)}}
    \sum_{k=-\infty}^\infty
    \exp\left[-\frac{(x-kd)^2}
    {2\sigma^2(1+i\hbar t/m\sigma^2)}\right].
    \end{split}
\end{equation}
With the use of definition of $\theta_3$-function (see, e.g. \cite{BE})
\begin{equation}\label{theta1}
    \theta_3(z,\tau)=\sum_{k=-\infty}^\infty\exp[i\pi(\tau k^2+2zk)]
\end{equation}
the wave function (\ref{inphase}) can be presented in the form
\begin{equation}\label{psi1}
    \psi(x,t)=\frac{\pi^{1/4}A}d\sqrt{\frac{2\sigma i}\tau}\,
    \exp\left({-\frac{i\pi x^2}{d^2\tau}}\right)
    \theta_3\left(\frac{x}{d\tau},-\frac1\tau\right),
\end{equation}
where
\begin{equation}\label{tau}
    \tau=\frac{2\pi i\sigma^2}{d^2}\left(1+\frac{i\hbar t}{m\sigma^2}
    \right).
\end{equation}
By means of transformation formula (see \cite{BE})
\begin{equation}\label{trans}
    \theta_3\left(\frac{z}\tau,-\frac1\tau\right)=\sqrt{-i\tau}\,
    \exp\left({\frac{i\pi z^2}\tau}\right)\theta_3(z,\tau)
\end{equation}
we transform (\ref{psi1}) into
\begin{equation}\label{psi2}
    \psi(x,t)=\frac{\sqrt{2\pi\sigma}\,A}{\pi^{1/4}d}\,\theta_3\left(
    \frac{x}d,\tau\right).
\end{equation}
Then the periodicity property of $\theta_3$-function,
$ \theta_3(z,\tau\pm 2)=\theta_3(z,\tau),$
leads at once to the periodicity of the wave function,
\begin{equation}\label{per2}
    \psi(x,t+t_r)=\psi(x,t),
\end{equation}
with the period
\begin{equation}\label{period}
    t_r=\frac{md^2}{\pi\hbar}.
\end{equation}

If the array of BECs is realized in optical periodic potential with
light wavelength $\lambda$, then the spacing between neighboring
lattice sites is equal to $d=\lambda/2$ and the revival time can
be expressed in the form
\begin{equation}\label{rev1}
    t_r=\frac14\cdot\frac{2\pi\hbar}{E_R},
\end{equation}
where
\begin{equation}\label{recoil}
    E_R=\frac{\hbar^2q^2}{2m}
\end{equation}
is the recoil energy ($q=2\pi/\lambda$).

The evolution time $t_r/2$ corresponds to the transformation of
$\theta_3$-function $\theta_3(z,\tau+1)=\theta_4(z,\tau)=
\theta_3(z+1/2,\tau)$, that is we obtain the wave function shifted
to the distance $d/2$ with respect to its initial form:
\begin{equation}\label{half}
    \psi(x,t+t_r/2)=\psi(x+d/2,t).
\end{equation}

\begin{figure}[ht]
\centerline{\includegraphics[width=10cm,height=8cm,clip]{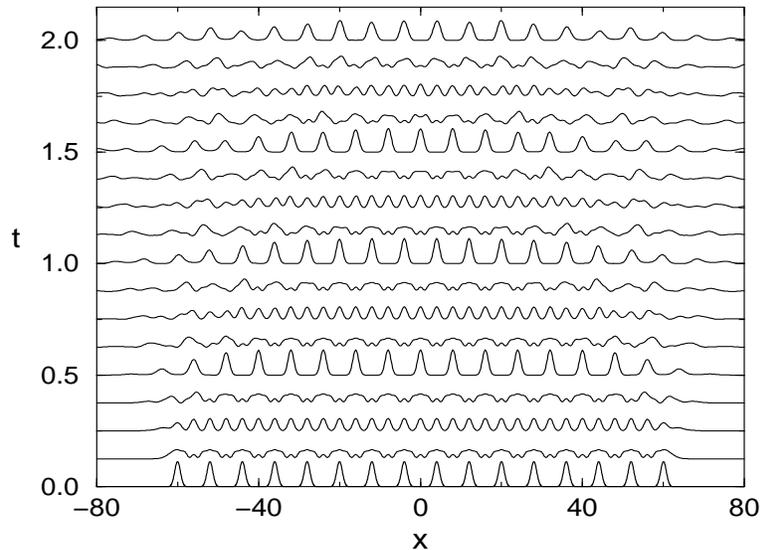}}
\vspace{0.3 true cm}
\caption{
Evolution of density profiles for BEC arrays with
zero relative phase. Two first revivals at $t=nt_r,$ $n=1,2,$
are clearly seen as well as ``fractional revivals'' at intermediate
moments $t_r/8,$ $t_r/4,$ $3t_r/8,$ $t_r/2,$ etc.
}
\label{fig2}
\end{figure}

Above calculation explains periodic restoration of initial wave
function by means of transformation properties of $\theta$-functions.
To relate this approach with standard one (see, e.g., \cite{AP2}),
let us consider the problem from a
different point of view. The linear Schr\"odinger equation
(\ref{FREE}) with periodic initial condition
\begin{equation}\label{initper}
\begin{split}
    \psi(x,0)=\frac{A}{\pi^{1/4}\sqrt{\sigma}}
    \sum_{k=-\infty}^\infty
    \exp\left[-\frac{(x-kd)^2}
    {2\sigma^2}\right]
    \end{split}
\end{equation}
can be solved by the Fourier method which gives
\begin{equation}\label{fourier}
    \psi(x,t)=\frac{\sqrt{2}\,\pi^{1/4}A}d\left\{1+2\sum_{k=1}^\infty
    \exp\left[-\frac{2\pi^2\sigma^2}{d^2}\left(1+\frac{i\hbar t}{m\sigma^2}
    \right)k^2\right]\cos\left(\frac{2\pi x}d k\right)\right\}.
\end{equation}
(The relationships between presentations (\ref{inphase}) and (\ref{fourier})
of the same wave function $\psi(x,t)$ is expressed by the known identity for these
two series; see, e.g., \cite{mumford}.)
We see that at $t$ equal to multiple of Talbot time $t_r$, $t=nt_r$, all phase
factors in (\ref{fourier}) become equal to unity and (\ref{fourier})
reduces to the Fourier series for the initial periodic wave function
(\ref{initper}). This method of derivation of time-periodicity of
the wave function shows that periodic restoration of the initial state is
not a specific feature of the initial state (\ref{initper}) built of
Gaussian functions. Indeed, any periodic initial function can be
expanded into Fourier series and harmonics $\cos(2\pi xk/d)$, $k=1,2,\ldots,$
evolve with time according to factors $\exp(-i\frac{2\pi^2\hbar t}{md^2}k^2)$
which become equal to unity at $t=nt_r$. Thus, any periodic initial wave
function completely restores periodically its form.
The described above picture of periodic in time changes of the
interference pattern is shown in Fig.~2 where even for relatively
small number of condensates first several revivals are clearly seen.

The above theory can be generalized on non-zero phases in the
initial state and hence in the solution (\ref{evolution}). For example,
in the case of alternating phases of condensates,
\begin{equation}\label{alt}
    e^{i\phi_k}=(-1)^k,
\end{equation}
the wave function can be expressed in terms of $\theta_4$-function
\cite{BE},
\begin{equation}\label{theta4}
\psi(x,t)=\frac{\pi^{1/4}A}d\sqrt{\frac{2\sigma i}\tau}\,
    \exp\left({-\frac{i\pi x^2}{d^2\tau}}\right)
    \theta_4\left(\frac{x}{d\tau},-\frac1\tau\right),
\end{equation}
or, with the use of the transformation formula \cite{BE},
\begin{equation}\label{trans2}
    \theta_4\left(\frac{z}\tau,-\frac1\tau\right)=\sqrt{-i\tau}\,
    \exp\left({\frac{i\pi z^2}\tau}\right)\theta_2(z,\tau),
\end{equation}
in the form
\begin{equation}\label{psi3}
    \psi(x,t)=\frac{\sqrt{2\pi\sigma}\,A}{\pi^{1/4}d}\,\theta_2\left(
    \frac{x}d,\tau\right).
\end{equation}
Then the property $\theta_2(z,\tau+1)=\exp(\pi i/4)\theta_2(z,\tau)$
leads to restoration of the initial state (up to inessential constant phase
factor) after revival time
\begin{equation}\label{rev3}
    t_r=\frac{md^2}{2\pi\hbar}=\frac18\cdot\frac{2\pi\hbar}{E_R}.
\end{equation}

Let us estimate an order of magnitude of the revival time for
arrays of BECs. In the case \cite{greiner2} of $^{87}$Rb
BECs array loaded into optical potential with light wavelength
$\lambda=838\,\mathrm{nm}$ formula (\ref{rev1}) gives
$t_r\simeq 75\,\mu\mathrm{s}$. This is about one order of
magnitude less than the revival time, caused by nonlinear interaction,
of single condensate in the experiment \cite{greiner2}.
In this experiment absorption images were taken after a
time-of-flight period of $16\,\mathrm{ms}$ which is much
greater (with factor $\sim 200$) than our estimate of $t_r$.
For number of sites in 3D lattice $\sim 10^3$ we have
$k_M\sim 10$ and, hence, the observed interference patterns correspond
to the Fraunhofer limit (\ref{cr4}). In this case the
difference of interference patterns was caused by difference
in initial states of condensates at different ``hold times''
of evolution of each condensate in strongly confined states formed
by 3D periodic trapping potential.

In the experiment \cite{Pedri} the  revival time is
$t_r\simeq 69\,\mu\mathrm{s}$ and a typical image was taken
at $t=29\,\mathrm{ms}$, that is for $k_M\sim100$ again in the Fraunhofer
limit (in accordance with the theory developed in this paper).

\section{Transition to Fraunhofer interference}

Now we shall turn to the regions (\ref{cr3}) and (\ref{cr4}).
Effects of Fresnel diffraction can be noticed in Fig.~1 for
$t=10t_r$. However, they are not expressed clearly enough
because of smooth distribution (\ref{Nk}) of density in the array
used in our calculations. Therefore it is more instructive to
consider finite array with equal amplitudes $A_k=1$ of wave
functions in each condensate and take $\phi_k=\Delta\phi\cdot k$,
that is with equal differences $\Delta\phi$ of phases between neighboring
condensates. Then Eq.~(\ref{train-time}) with $t\gg m\sigma^2/\hbar$
reduces to
\begin{equation}\label{4-1}
    \psi(x,t)\cong\frac1{\pi^{1/4}}\sqrt{\frac{m\sigma}{i\hbar t}}
    e^{\frac{imx^2}{2\hbar t}}e^{-\frac{m\sigma^2x^2}{2\hbar^2t^2}}
    \sum_{k=-k_M}^{k_M}\exp\left[-i\left(\frac{mdx}{\hbar t}
    -\Delta\phi\right)k+\frac{imd^2}{2\hbar t}k^2\right],
\end{equation}
where we have taken into account only leading real and imaginary
contributions in the series expansion of the exponential in powers
of $m\sigma^2/\hbar t$. The sum here has maximal amplitude when all
terms are in phase in linear in $k$ approximation. This condition
defines coordinates $x_n$ of the centers of fringes,
\begin{equation}\label{4-2}
    x_n=\frac{2\pi\hbar}{md}\left(n+\frac{\Delta\phi}{2\pi}\right)t,
    \quad n=0,\pm1,\pm2,\ldots
\end{equation}

To consider profiles of fringes, we introduce the coordinate $\delta$
which is reckoned from the center of the fringe:
\begin{equation}\label{4-3a}
    x=x_n+\delta,
\end{equation}
so that dependence on $\delta$ is determined mainly by the factor
\begin{equation}\label{4-3}
    \Phi(\delta,t)=\sum_{k=-k_M}^{k_M}\exp\left(-\frac{imd\delta}{\hbar t}k
    +\frac{imd^2}{2\hbar t}k^2\right).
\end{equation}
If $t$ satisfies the condition (\ref{cr3}), then both terms in the
exponential have the same order of magnitude and, on one hand, the
fringe width is of order of magnitude of the array length,
$\delta\sim 2k_Md$, and, on the other hand, it is much less than
the distance between fringes. Therefore the coordinate $x$ in the
factor $\exp\left(-{m^2\sigma^2x^2}/{2\hbar^2t^2}\right)$ can be
replaced by $x_n$. Thus, the wave function in vicinity of the $n$th
fringe is given by
\begin{equation}\label{4-4}
    \psi_n(x,t)=\frac1{\pi^{1/4}}\sqrt{\frac{m\sigma}{i\hbar t}}\,
    e^{\frac{imx^2}{2\hbar t}}\cdot\exp\left[-\frac{2\pi^2\sigma^2}
    {d^2}\left(n+\frac{\Delta\phi}{2\pi}\right)^2\right]\,
    \Phi(\delta,t),
\end{equation}
where $\delta=x-x_n$. Now, for $k_M\gg1$ the sum in (\ref{4-3})
can be approximated by integrals which are easily expressed
in terms of Fresnel functions \cite{GR}:
\begin{equation}\label{4-5}
\begin{split}
    \Phi(\delta,t)=\sqrt{\frac{\pi\hbar t}{md^2}}e^{\frac{im\delta^2}{2\hbar t}}
    &\Bigg[\fC\left(\sqrt{\frac{m}{2\hbar t}}(k_Md+\delta)\right)+
    \fC\left(\sqrt{\frac{m}{2\hbar t}}(k_Md-\delta)\right)\\
    &+i\left(
    \fS\left(\sqrt{\frac{m}{2\hbar t}}(k_Md+\delta)\right)+
    \fS\left(\sqrt{\frac{m}{2\hbar t}}(k_Md-\delta)\right)\right)\Bigg].
    \end{split}
\end{equation}
Thus, distribution of density in the $n$th fringe is given by
\begin{equation}\label{4-6}
\begin{split}
    |\psi_n|^2=\frac{\sqrt{\pi}\sigma}{d^2}&\exp\left[-\frac{4\pi^2\sigma^2}
    {d^2}\left(n+\frac{\Delta\phi}{2\pi}\right)^2\right]\\
    &\times\Bigg\{\left[\fC\left(\sqrt{\frac{m}{2\hbar t}}(k_Md+\delta)\right)+
    \fC\left(\sqrt{\frac{m}{2\hbar t}}(k_Md-\delta)\right)\right]^2\\
    &+\left[\fS\left(\sqrt{\frac{m}{2\hbar t}}(k_Md+\delta)\right)+
    \fS\left(\sqrt{\frac{m}{2\hbar t}}(k_Md-\delta)\right)\right]^2\Bigg\}.
    \end{split}
\end{equation}
The exponential factor determines the number of atoms in the $n$th fringe:
\begin{equation}\label{4-7}
    N_n=\mathrm{const}\cdot\exp\left[-\frac{4\pi^2\sigma^2}
    {d^2}\left(n+\frac{\Delta\phi}{2\pi}\right)^2\right].
\end{equation}
This formula reduces to Eq.~(6) of Ref.~\cite{Pedri} for $\Delta\phi=0$.
\begin{figure}[ht]
\centerline{\includegraphics[width=8cm,height=12cm,clip]{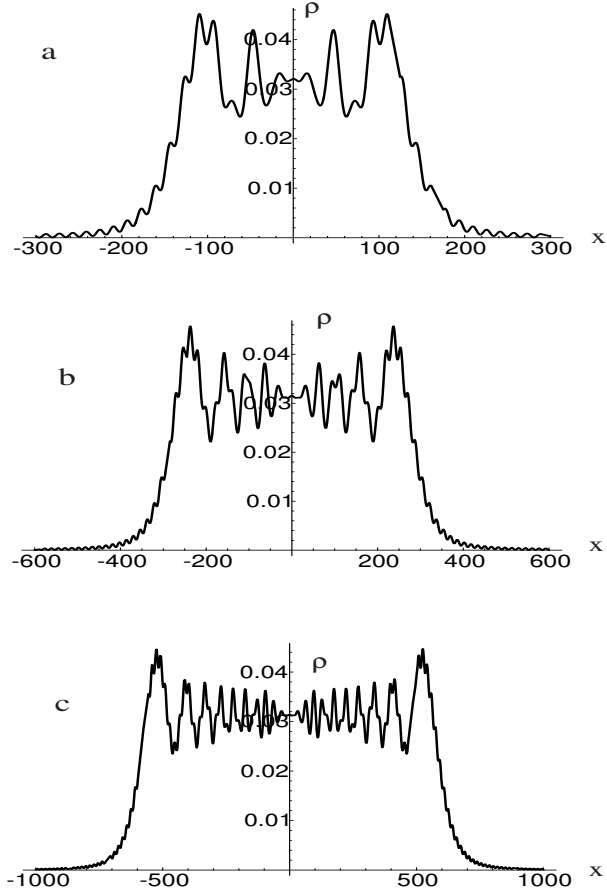}}
\vspace{0.3 true cm}
\caption{
The central fringe profile for several values of the number of sites
in the array. Time $t$ corresponds to the region (\ref{cr3}).
The plots are calculated
for $d=8$, $\sigma=1$ and (a) $k_M=20$ at $t=40t_r$;
(b) $k_M=40$ at $t=80t_r$; (c) $k_M=80$ at $t=160t_r$.
Formation of the Fresnel pattern is clearly seen.
}
\label{fig3}
\end{figure}

Dependence on $\delta$ determines fine interference pattern inside fringes.
It is expressed by the factor in curly brackets and demonstrates typical
Fresnel form (see, e.g., \cite{FH65}, section 3.3, or \cite{BW}, section 8.7)
of diffraction from a slit with width $2k_Md$ equal to the whole array
length. Accuracy of this analytical description depends on the number of
sites in the array and increases with growth of $k_M$. In Fig.~3 it is shown
how exact profile of density along the fringe changes with increase
of $k_M$. Its transformation into Fresnel diffraction pattern is clearly seen.
Small ``ripples'' are obviously caused by the discrete structure of the
array.

\begin{figure}[ht]
\centerline{\includegraphics[width=12cm,height=8cm,clip]{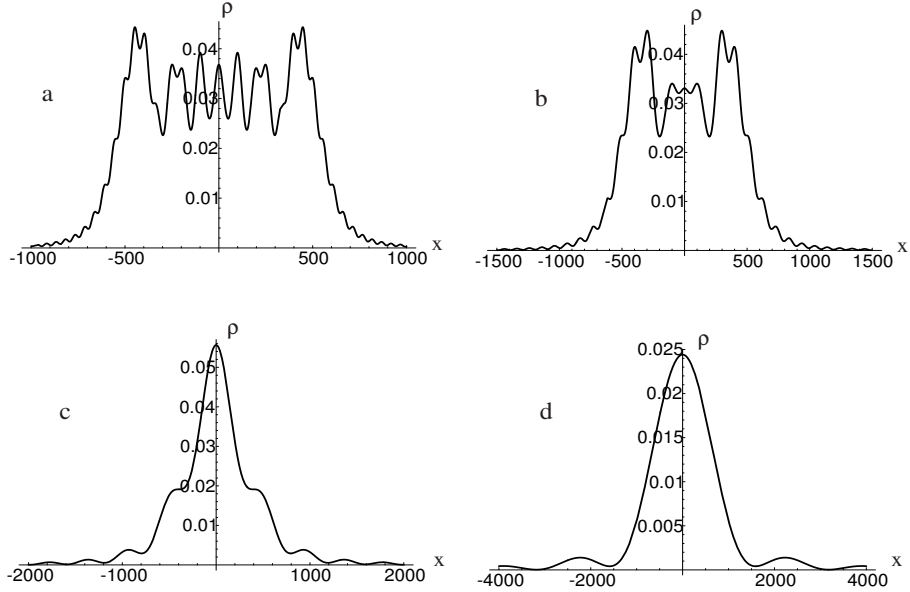}}
\vspace{0.3 true cm}
\caption{
Evolution of density profile of central fringe on time. Values of the
parameters are equal to $d=8$, $\sigma=1$, $k_M=80$ and (a) $t=500t_r$;
(b) $t=1000t_r$; (c) $t=2000t_r$; (d) $t=16000t_r$. Transformation of
Fresnel profile shown in Fig.~3~(c) to standard Fraunhofer profile is
clearly seen.
}
\label{fig4}
\end{figure}

For larger values of time (\ref{cr4}) Fresnel structure evolves into
usual form of density distribution in Fraunhofer diffraction from
finite slit with width $2k_Md$. In this limit of very large $t$
the quadratic in $k$ tern in exponentials in Eq.~(\ref{4-3}) is
much less than unity and can be omitted. Then simple integration gives
$\Phi(\delta,t)=\frac{2\hbar t}{nd\delta}\sin\left(\frac{mk_Md\delta}
{\hbar t}\right)$ and hence distribution of density inside fringes
is proportional to
\begin{equation}\label{4-8}
    |\Phi(\delta,t)|^2=4\frac{\sin^2(mk_Md\delta/\hbar t)}{(md\delta/
    \hbar t)^2}
\end{equation}
which is standard Fraunhofer diffraction distribution from finite
slit (see, e.g., \cite{BW}, section 8.5).
The described here evolution of profile is illustrated in Fig.~4.
The total intensity of $n$th fringe
is still determined, of course, by Eq.~(\ref{4-7}).

\section{Conclusion}

We have presented in this paper analysis of interference of matter waves
during one-dimensional expansion of finite arrays of condensates. It shows
that the interference pattern exhibits quite complicated  evolution
with time from Talbot ``collapses and revivals'' of wave function
through intermediate region of Fraunhofer fringes with Fresnel
patterns inside them, and, eventually, to standard Fraunhofer
diffraction from finite grating. One may suppose that technique
of density imaging will permit one to study experimentally all
these stages.

\subsection*{Acknowledgements}
This work was supported by FAPESP (Brazil) and CNPq (Brazil).
A.M.K. thanks also RFBR for partial support.

\end{document}